\documentclass[a4paper,fleqn]{cas-sc}
\usepackage{aliascnt}
\usepackage[numbers]{natbib}

\def\tsc#1{\csdef{#1}{\textsc{\lowercase{#1}}\xspace}}
\tsc{WGM}
\tsc{QE}
\tsc{EP}
\tsc{PMS}
\tsc{BEC}
\tsc{DE}

\begin{document}
\let\WriteBookmarks\relax
\def\floatpagepagefraction{1}
\def\textpagefraction{.001}

\shorttitle{Targeted influence maximization}
 
\shortauthors{R.Zhang et~al.}

\title [mode = title]{Targeted influence maximization in complex networks}                      



%
\author[1]{Renquan Zhang}[style=chinese,
                        orcid=0000-0002-4927-1827]




\credit{Designed this study, Wrote the code and first draft of the manuscript}

\address[1]{School of Mathematical Sciences, Dalian University of Technology,Dalian,116024,China}

\author[1]{Xiaolin Wang}[style=chinese]

\credit{Wrote the code, ran simulations and performed the analysis}

\author[2]{Sen Pei}[style=chinese,%
   orcid=0000-0002-7072-2995
   ]
\cormark[1]

\credit{Designed this study, reviewed and edited the manuscript}

\address[2]{Department of Environmental Health Sciences, Mailman School of Public Health, Columbia University,New York,10032,NY,USA}

\cortext[cor1]{Corresponding author}


\begin{abstract}
Many real-world applications based on spreading processes in complex networks aim to deliver information to specific target nodes. However, it remains challenging to optimally select a set of spreaders to initiate the spreading process. In this paper, we study the targeted influence maximization problem using a susceptible-infected-recovered (SIR) model as an example. Formulated as a combinatorial optimization, the objective is to identify a given number of spreaders that can maximize the influence over target nodes while minimize the influence over non-target nodes. To find a practical solution to this optimization problem, we develop a theoretical framework based on a message passing process and perform a stability analysis on the equilibrium solution using non-backtracking (NB) matrices. We propose that the spreaders can be selected by imposing optimal perturbation on the equilibrium solution for the subgraph consisting of the target nodes and their multi-step nearest neighbors while avoiding such perturbation on the complement graph that excludes target nodes from the original network. We further introduce a metric, termed targeted collective influence, for each node to identify influential spreaders for targeted spreading processes. The proposed method, validated in both synthetic and real-world networks, outperforms other competing heuristic approaches. Our results provide a framework for analyzing the targeted influence maximization problem and a practical method to identify spreaders in real-world applications.

\end{abstract}


\begin{highlights}
\item We developed a theoretical framework to analyze targeted influence maximization using a message passing process.
\item We proposed a metric based on non-backtracking matrix to select influential spreaders.
\item We validated the proposed metric in both synthetic and real-world networks.
\end{highlights}

\begin{keywords}
Influential Spreaders \sep Targeted Influence Maximization \sep Spreading Dynamics \sep Targeted Spreading \sep Non-backtracking Matrix \sep Optimal Perturbation
\end{keywords}

\maketitle

\section{Introduction}

Spreading processes in complex networks can describe a wide variety of real-world phenomena, ranging over epidemic outbreaks \cite{pastor2015epidemic,newman2002spread,pastor2001epidemic}, online information diffusion \cite{zhang2016dynamics,watts2007influentials,goel2016structural,zhou2020realistic}, behavior adoption \cite{centola2010spread,granovetter1978threshold,aral2011creating}, and viral marketing\cite{domingos2001mining,leskovec2007dynamics}. Due to the structural heterogeneity of networks, a small set of nodes play a disproportionate role in shaping the outcome of spreading dynamics. Identifying such pivotal nodes, or influencers, is a critical question in network science. Over the last decades, a plethora of studies developed methods to locate influential spreaders in networks \cite{kitsak2010identification,aral2012identifying,pei2018theories,lu2016vital}, either a single node initiating the spreading process or multiple spreaders considering their collective influence \cite{kempe2003maximizing,chen2009efficient,morone2015influence,pei2014searching,aral2018social,teng2016collective,leskovec2007cost,braunstein2016network,pei2013spreading,clusella2016immunization,radicchi2016leveraging,ren2019generalized}. A review on recent advances in this area can be found in Ref \cite{pei2020influencer}.

Many real-world applications aim to deliver information to certain target nodes while avoiding reaching non-target nodes. For instance, in advertising, it is desirable to promote products to potential customers and minimize the coverage among other users for precise marketing; in an election campaign, it is more cost-effective to disseminate information to swing voters and save the resources spent on decided voters. Studies on targeted spreading and control have attracted much attention in recent years \cite{gao2014target,cornelius2013realistic,song2016targeted,li2015real,calio2018topology,calio2021attribute,ke2018finding,su2018location,sun2016spreading}. A number of heuristic methods were developed to identify influencers for targeted spreading processes. For instance, a greedy algorithm was proposed to select seeds in a spreading process with several constrains \cite{song2016targeted}; a heuristic method based on local path counting was used to identify single spreaders in a targeted spreading process \cite{sun2016spreading}; and real-time targeted online advertisement informed by key words was also tested \cite{li2015real}. While these approaches were demonstrated effective in different settings, a general framework for analyzing the targeted influence maximization problem is lacking.

In this study, we focus on the influence maximization problem for a general targeted spreading process. Specifically, we aim to identify a given number of spreaders that maximize the influence over target nodes and minimize the influence over non-target nodes in a susceptible-infected-recovered (SIR) model. To solve this optimization problem, we first develop a mathematical framework to formulate the system as a message passing process, and then perform a stability analysis on the equilibrium solution using the non-backtracking (NB) matrix of the system \cite{hashimoto1989zeta}. We argue that the spreaders can be selected by imposing optimal perturbation on the equilibrium solution for the subgraph consisting of the target nodes and their multi-step nearest neighbors while avoiding such perturbation on the complement graph that excludes target nodes from the original network. Our analysis leads to a theoretically based metric, termed targeted collective influence, to quantify the targeted influence of each node, which allows the selection of multiple influencers in the targeted spreading process. We validate the proposed method in both synthetic and real-world networks and demonstrate that it outperforms commonly used heuristic approaches. Our analysis provides a theoretical framework for analyzing the targeted influence maximization problem and a practical method to identify spreaders in real-world applications.

\section{Model}

We use a susceptible-infectious-recovered (SIR) agent-based model to simulate the spreading process in networks. Considering a network composed of $N$ nodes and $M$ undirected edges, denote $\{a_{ij}\}_{N\times N}$ as the binary adjacency matrix ($a_{ij}=1$ if node $i$ is connected to node $j$, and $a_{ij}=0$ otherwise). The binary variables $S_i(t)$, $I_i(t)$ and $R_i(t)$ represent that node $i$'s state is susceptible, infectious and recovered at time $t$, respectively. We denote the probability that an infectious node will infect its susceptible neighbor as $\beta$, and define $\gamma$ as the infectious period (without loss of generality, $\gamma=1$). At each time step $t$, a susceptible node $i$ ($S_i(t)=1$) can be infected by its neighbor $j$ in the infectious state ($I_j(t)=1$) with probability $\beta$. Meanwhile, nodes in the infectious state will transit to the recovered state after $\gamma$ steps and can never be infected again. The spreading process can be described as follows: 
\begin{align}
&\frac{d S_i(t)}{dt}=-S_i(t)\big[1-\prod_{j}(1-\beta a_{ij}I_j(t))\big],\label{sir1}\\ 
&\frac{d I_i(t)}{dt}=S_i(t)\big[1-\prod_{j}(1-\beta a_{ij}I_j(t))\big]-\frac{I_i(t)}{\gamma}, \label{sir2}\\ 
&\frac{d R_i(t)}{dt}=\frac{I_i(t)}{\gamma}. \label{sir3}
\end{align}

Here we focus on a combinatorial optimization problem - how to select a given number of initial infected nodes, or seeds, to maximize infection among a specific group of nodes and minimize infection among others? Specifically, denote $V=V^T\bigcup V^{NT}$ as the set of nodes, where $V^T$ represents the set of target nodes and $V^{NT}$ the set of non-target ones. The seeds can be only selected from $V^{NT}$ and the number of seeds is denoted as $n^*$. Based on the SIR dynamics, the number of nodes that have been infected is equal to the number of the recovered nodes at the end of the spreading process. For a given network, the influence over the target and non-target nodes are defined as 
\begin{align}
&f(s_{n^*})=\lim_{t\to\infty}\sum_{i\in V^T} R_i(t), \label{goal1} \\ 
&g(s_{n^*})=\lim_{t\to\infty}\sum_{i\in V^{NT}} R_i(t), \label{goal2}
\end{align}
where $s_{n^*}$ denotes the set of seeds with the size $n^*$. As the topological structure of the network plays a significant role in the spreading process, different combinations of seeds with the same size $n^*$ could lead to contrasting outcomes. We aim to find the optimal set of seeds such that $g(s_{n^*})/f(s_{n^*})$ is minimized for $f(s_{n^*})>0$.

\section{Method}

\subsection{Message passing equations}

We first formulate the propagation as a message passing process. Compared to the master equations defined using the adjacency matrix, the message passing process can better represent the SIR dynamics. Specifically, in the SIR model, backtracking infections ($i\to j\to i$) are not allowed as the transmission is irreversible. The adjacency matrix allows backtracking infections, which can introduce excessive dynamical resonance between pairs of connected nodes. In contract, the message passing process excludes backtracking spreading and was found superior in analyzing a number of dynamical models in complex networks \cite{morone2015influence,pei2017efficient,karrer2014percolation,hamilton2014tight,wang2018optimal,aleja2019non,zhang2018dynamic,wang2019stability,martin2014localization,zhang2020backtracking,kawamoto2016localized,bordenave2015non}.

To study the impact of a node $j$ on its neighbor node $i$, we investigate the probability of node $i$ being infected if node $j$ is assumed to be absent from the network. For a link from $i$ to $j$ ($i\to j$, even if the link is undirected), suppose node $j$ is "virtually" removed from the network (i.e., creating a "cavity" at node $j$) and calculate the probability of node $i$ being infected in the absence of node $j$ at time $t$, which is represented as $S_{i\to j}(t)$. We apply the same procedure for $I$ and $R$. For sparse networks without too many short loops, the message passing process can be described by
\begin{align}
&S_{i\to j}(t+1)=S_{i\to j}(t)\prod_{k\backslash j}\left(1-\beta a_{ik}I_{k\to i}(t)\right), \label{mp1}\\ 
&I_{i\to j}(t+1)=S_{i\to j}(t)\big[1-\prod_{k\backslash j}\left(1-\beta a_{ik}I_{k\to i}(t)\right)\big]+I_{i\to j}(t)(1-\frac{1}{\gamma}), \label{mp2}\\
&R_{i\to j}(t+1)=R_{i\to j}(t)+\frac{I_{i\to j}(t)}{\gamma}. \label{mp3}
\end{align}
Here $k\backslash j$ means $k$ runs over all nodes except $j$. Denote $\lim\limits_{t\to \infty} S_{i\to j}(t)=S_{i\to j}$, $\lim\limits_{t\to \infty} I_{i\to j}(t)=I_{i\to j}$ and $\lim\limits_{t\to \infty} R_{i\to j}(t)=R_{i\to j}$. Note that Eq. (\ref{mp3}) is redundant. The steady state of the nonlinear dynamical system can be obtained by solving the following self-satisfying equations:
\begin{align}
&S_{i\to j}=S_{i\to j}\prod_{k\backslash j}\left(1-\beta a_{ik}I_{k\to i}\right), \label{steady1}\\ 
&I_{i\to j}=\gamma S_{i\to j}\big[1-\prod_{k\backslash j}\left(1-\beta a_{ik}I_{k\to i}\right)\big]. \label{steady2}
\end{align}

\begin{figure}
	\centering
		\includegraphics[scale=.45]{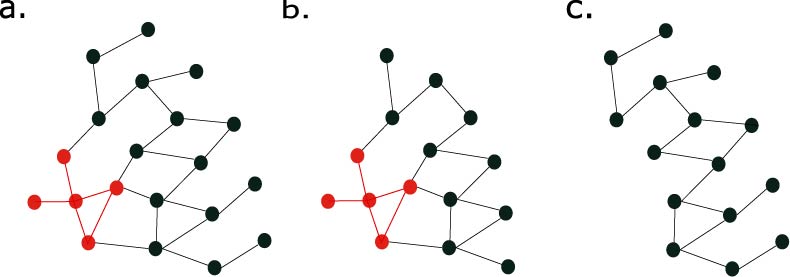}
	\caption{Illustration of the subgraph $G_s$ and $G_c$. a, The original network with $N=20$ and 5 target nodes highlighted in red. b, The subgraph $G_s$ with $L=2$. c, The subgraph $G_c$. The candidate seeds are the black nodes in b.}
	\label{FIG:0}
\end{figure}

\subsection{Stability analysis}

To maximize influence over the target nodes and minimize influence over the non-target nodes, we consider two subgraphs of the original network: 1) a subgraph $G_s$ consisting of the target nodes and their nearest neighbors within $L$ steps, and 2) a subgraph $G_c$ that excludes the target nodes from the original network. We create $G_s$ using a breadth-first-search algorithm starting from the target nodes. An illustration for $G_s$ ($L=2$) and $G_c$ is shown in Fig. \ref{FIG:0}. As seeds can be only selected from the non-target nodes, we define the set of candidate seeds as the non-target nodes in $G_s$ (i.e., nodes in $G_s$ excluding the target nodes).

For both $G_s$ and $G_c$, a trivial equilibrium solution exists: $(S_{i\to j}^*,I_{i\to j}^*)^T=(\mathbf{1},\mathbf{0})^T$, corresponding to the state that all nodes are susceptible. The stability of the trivial solution is controlled by the largest eigenvalue of the Jacobian matrix ($\mathbf{J}$) at this solution. Now we derive the Jacobian matrix at the solution $(\mathbf{1},\mathbf{0})^T$ for $G_s$ and $G_c$. For a given network, we take the partial derivatives of Eq. (\ref{steady1}). For the  directed links $k\to l$ and $i\to j$, we have
\begin{equation}\label{p-ss}
\frac{\partial S_{i\to j}}{\partial S_{k\to l}}=\prod_{k\backslash j}\left(1-\beta a_{ik}I_{k\to i}(t)\right)\big\vert_{(\mathbf{1},\mathbf{0})}=\left\{
\begin{aligned}
&1 &  &\mbox{if}\; k=i\;\mbox{and}\; l=j\\ 
&0      &  &\mbox{otherwise}
\end{aligned}
\right.
\end{equation}
\begin{equation}\label{p-si}
\frac{\partial S_{i\to j}}{\partial I_{k\to l}}=-\beta S_{i\to j}(t)\prod_{k^{'}\backslash j,k}\left(1-\beta a_{ik^{'}}I_{k^{'}\to i}(t)\right)\bigg\vert_{(\mathbf{1},\mathbf{0})}=\left\{
\begin{aligned}
&-\beta &  &\mbox{if}\; l=i \;\mbox{and}\; k\neq j\\ 
&0      &  &\mbox{otherwise}
\end{aligned}
\right.
\end{equation}
The same analysis on Eq. (\ref{steady2}) yields:
\begin{align}\label{p-is}
\frac{\partial I_{i\to j}}{\partial S_{k\to l}}=\gamma\big[1-\prod_{k\backslash j}(1-\beta a_{ik}I_{k\to i}(t))\big]\big\vert_{(\mathbf{1},\mathbf{0})}=0
\end{align}
\begin{equation}\label{p-ii}
\frac{\partial I_{i\to j}}{\partial I_{k\to l}}=\beta\gamma S_{i\to j}(t)\big[\prod_{k^{'}\backslash j,k}(1-\beta a_{ik^{'}}I_{k^{'}\to i}(t))\big]\bigg\vert_{(\mathbf{1},\mathbf{0})}=\left\{
\begin{aligned}
&\beta\gamma &  &\mbox{if}\; l=i \;\mbox{and}\; k\neq j\\ 
&0     &  &\mbox{otherwise}
\end{aligned}
\right.
\end{equation}
So the Jacobian matrix at the solution $(\mathbf{1},\mathbf{0})^T$ is given by:
\begin{equation}\label{jacpbi}
\mathbf{J}\bigg\vert_{(\mathbf{1},\mathbf{0})}=\left(
\begin{aligned}
\mathbf{I}&   &\mathbf{A} \\ 
\mathbf{0}&   &\mathbf{B}
\end{aligned}
\right).
\end{equation}
Here $\mathbf{I}$ is the identity matrix, $\mathbf{A}=\left\{ \frac{\partial S_{i\to j}}{\partial I_{k\to l}}\right\}_{2M\times 2M}$ and $\mathbf{B}=\left\{ \frac{\partial I_{i\to j}}{\partial I_{k\to l}}\right\}_{2M\times 2M}$, where $M$ is the number of links. The stability of $I_{i\to j}^*$ is determined by the largest eigenvalue of the matrix $\mathbf{B}$, denoted by $\lambda_{B}$. The trivial solution is stable if $\lambda_{B}<1$ and unstable if $\lambda_{B}>1$. The matrix $\mathbf{B}$ is a generalization of the non-backtracking (NB) matrix of the network $\mathbf{W}$, which was found important for a range of dynamical processes in complex networks. Precisely, $\mathbf{B}=\beta\gamma\mathbf{W}$, where 
\begin{equation}\label{NB}
\mathbf{W}_{k\to l,i\to j}=\left\{
\begin{aligned}
&1     &  &\mbox{if}\; l=i \;\mbox{and}\; k\neq j,\\ 
&0     &  &\mbox{otherwise.}
\end{aligned}
\right.
\end{equation}

\subsection{Optimal perturbation}

Selecting seeds to initiate a spreading process acts as a perturbation on the equilibrium solution. Define $\mathbf{I}_\to(0)=(\cdots,I_{i\to j}(0),\cdots)^T$ as the initial state. For the equilibrium solution, we have $\mathbf{I}_\to(0)=\mathbf{0}^T$. If node $i$ is chosen as a seed, we set the elements $I_{i\to j}(0)=1$ for all $j\in\partial i$, where $\partial i$ represents the set of neighbors of node $i$. To select a given number of $n^*$ seeds, there exist a total of $C^N_{n^*}$ possible combinations. In order to maximize influence over the target nodes, we can impose the optimal perturbation along the leading eigenvector of the NB matrix $\mathbf{W}$ for $G_s$ so that more nodes are infected in this subgraph. Meanwhile, we should avoid the perturbation in $G_c$ along the leading eigenvector of its NB matrix to minimize infections among non-target nodes. Similar approaches have been used in numerical weather prediction \cite{Palmer_2000,toth1993ensemble,toth1997ensemble} and infectious disease forecasting \cite{pei2019predictability,pei2017counteracting,pei2021optimizing}.


Denote the leading eigenvector of $\mathbf{W}$ as $\mathbf{v}$ such that $\mathbf{Wv}=\lambda\mathbf{v}$, where $\lambda$ is the largest eigenvalue of $\mathbf{W}$. Computing the leading eigenvector for $\mathbf{W}$ can be challenging for large-scale networks due to the high dimensionality of the NB matrix. For a network with $M$ edges, $\mathbf{W}$ has a dimension of $2M\times 2M$. An effective method to compute the largest eigenvalue and the leading eigenvector is the power iteration. Starting from an initial vector $\mathbf{y}_0=\mathbf{1}^T$, we multiple $\mathbf{W}$ from the left ($\mathbf{y}_{t+1}=\mathbf{Wy}_t$) repeatedly until the ratio $\|\mathbf{y}_{t+1}\|/\|\mathbf{y}_t\|$ is stabilized. The largest eigenvalue is $\lambda=\lim_{t\to\infty}\|\mathbf{y}_{t+1}\|/\|\mathbf{y}_t\|$ and the normalized leading eigenvector is $\mathbf{v}=\lim_{t\to\infty}\mathbf{y}_t/\|\mathbf{y}_t\|$.


To approximate the leading eigenvector using the power iteration, we multiple $\mathbf{W}$ from the left on $\mathbf{y}_0=\mathbf{1}^T$ for $\ell$ times. We denote the approximated influence of node $i$ for $\ell$ as $CI_\ell(i)=\sum_{j\in\partial i}y_{\ell, i\to j}^2$, where $y_{\ell, i\to j}$ is the entry of $\mathbf{y}_\ell$ corresponding to the link $i\to j$. Following the method in Ref. \cite{morone2015influence}, we derive that $CI_\ell(i)$ for node $i$ in a network is given by
\begin{align}
CI_{\ell}(i)=\left(k_{i}-1\right)\sum_{j\in \partial Ball(i,2\ell-1)}\left(k_{j}-1\right), \label{ci}
\end{align}
where $\ell$ is the iteration time and $\partial Ball(i,2\ell-1)$ is the set of nodes whose shortest distance to node $i$ is $2\ell-1$. In order to find spreaders for the targeted spreading process, we define the targeted collective influence for node $i$ at level $\ell$ as 
\begin{align}
\Delta CI_{\ell}(i)=CI_{\ell}^{G_{s}}(i)-CI_{\ell}^{G_{c}}(i). \label{VS}
\end{align}
Here $CI_{\ell}^{G_{s}}(i)$ and $CI_{\ell}^{G_{c}}(i)$ are calculated on the subgraph $G_{s}$ and $G_{c}$. For the targeted influence maximization problem, we select top $n^{*}$ nodes from the candidates with the highest $\Delta CI_{\ell}$ score as the seed set $s_{n^*}$. Nodes with higher $\Delta CI_{\ell}$ scores tend to have higher $CI_{\ell}^{G_{s}}$ and lower $CI_{\ell}^{G_{c}}$.

\section{Numerical validation}

In order to test the performance of the proposed targeted collective influence, we first run numerical simulations on synthetic networks. We consider the case that target nodes are connected in a local cluster. In experiments, we first randomly select a target node and then apply a breadth-first search to assign other target nodes until the predefined number of target nodes is reached. Without loss of generality, here we set 10 target nodes in each cluster. Once the target node set $V^{T}$ is assigned, we select spreaders using different methods. Starting from the selected seeds, we perform 100 independent realizations of the SIR model. Results are evaluated using the average of the 100 simulations.

To compute the targeted collective influence $\Delta CI_\ell$, we define $G_s$ as the subgraph consisting of the target nodes and their nearest neighbors within $L=7$ steps. Other values of $L$ were tested. We find that $L=7$ is enough to capture potential optimal spreaders and increasing $L$ does not improve the performance. In model simulations, we consider $\ell=1$ and $\ell=2$ to calculate $\Delta CI_\ell$. We compare the targeted collective influence $\Delta CI_1$ and $\Delta CI_2$ with several other competing methods, including (1) dynamical importance defined based on the eigenvector of the adjacency matrix of the original network $G$ (Eig) \cite{restrepo2006characterizing}; (2) the dynamical importance defined based on the eigenvector of the adjacency matrix of the subgraph $G_s$ (Eigs); (3) the degree centrality of $G$ (HD); (4) the degree centrality of $G_s$ (HDs); (5) the collective influence $CI_\ell$ of $G$ for $\ell=1$ in Eq. (\ref{ci}) ($CI_1$); and (6) the collective influence $CI_\ell$ of $G$ for $\ell=2$ in Eq. (\ref{ci}) ($CI_2$). More details of the competing methods are provided in Appendix \ref{othermethod}. For each metric, we select the top $n^*$ nodes with the highest values as the initial seeds. For reference, we also test a random selection method (Rand) that chooses seeds randomly from $G$.

\begin{figure}
	\centering
		\includegraphics[scale=.5]{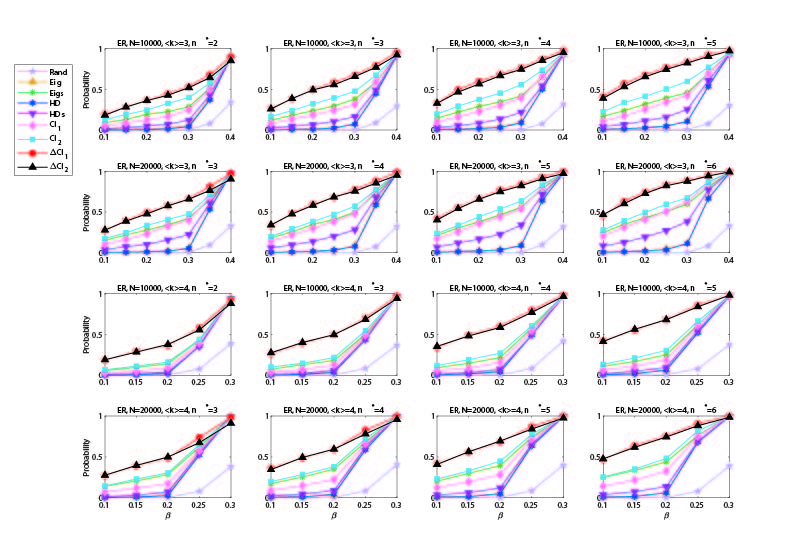}
	\caption{The probability that at least one target node is infected in Erd\"{o}s-R\'{e}nyi (ER) networks with size $N=10,000$ or $20,000$ and mean degree $\left<k\right>=3$ or $4$. We use different methods (including Rand, Eig, Eigs, HD, HDs, CI and $\Delta CI$ with $\ell=1$ and $\ell=2$) to select seeds for different size $n^{*}=2,\cdots,6$. The results are averaged over 100 independent realizations.}
	\label{FIG:1}
\end{figure}

\begin{figure}
	\centering
		\includegraphics[scale=.5]{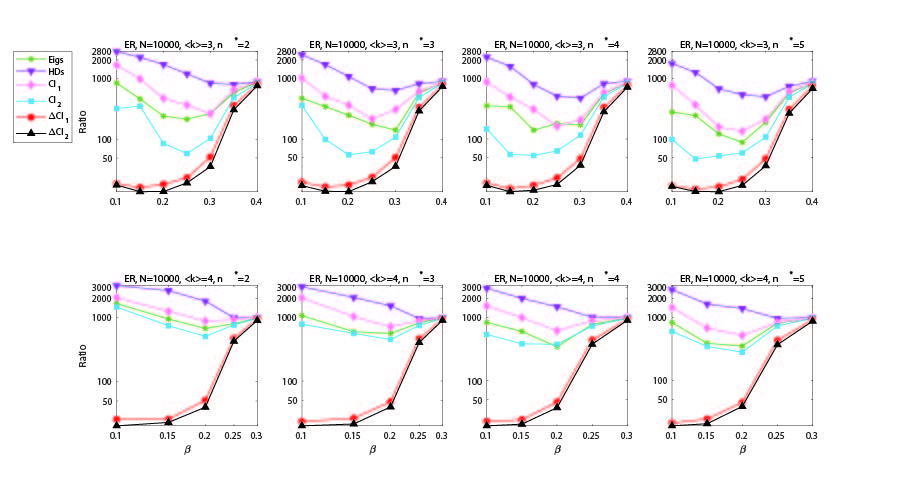}
	\caption{The ratio of $g(s_{n^*})$ to $f(s_{n^*})$ for Erd\"{o}s-R\'{e}nyi (ER) networks with size $N=10,000$ and mean degree $\left<k\right>=3$ or $4$. We compare the performance of different methods including Eigs, HDs, CI and $\Delta CI$ with $\ell=1$ and $\ell=2$. The results are averaged over 100 independent realizations.}
	\label{FIG:2}
\end{figure}

\subsection{Random networks}

We first test on homogeneous Erd\"{o}s-R\'{e}nyi (ER) random networks with 10 target nodes. We generate undirected ER networks with size $N$ and average degree $\left<k\right>$ by randomly connecting any possible pairs of nodes with a probability $p=\left<k\right>/N$. We use networks with $N=10,000$ or $20,000$ and $\left<k\right>=3$ or $4$ in simulations shown in Figs. \ref{FIG:1} and \ref{FIG:2}. To ensure the connectivity of the graph, all simulations are only applied on the giant connected component. We vary the transmission rate $\beta\in \left[0.1,0.4\right]$ for $\left<k\right>=3$ and $\beta\in \left[0.1,0.3\right]$ for $\left<k\right>=4$. We don't consider higher transmission rate $\beta$ as it will lead to large-scale outbreaks that infect almost the entire network. We test the number of seeds $n^*=2,\cdots,6$ and set $\gamma=1$.

In Fig. \ref{FIG:1}, we show the probability that at least one target node is infected, $P_t$. As all target nodes are locally connected (as shown in Fig. \ref{FIG:0}a), $P_t$ measures the chance that the spreading process reaches the small cluster of target nodes in a large-scale network. The targeted collective influence $\Delta CI_{\ell}$ consistently outperforms other competing methods. However, $\Delta CI_1$ and $\Delta CI_2$ have similar results. As the transmission rate $\beta$ increases, $P_t$ increases for all methods. For $\beta<\langle k\rangle/(\langle k^2\rangle-\langle k\rangle)$, seeds selected by Rand, Eig, and HD can hardly infect any target nodes. For $\beta>\langle k\rangle/(\langle k^2\rangle-\langle k\rangle)$, all methods except Rand can almost always reach target nodes. Metrics defined on the subgraph $G_s$ performs better than their counterparts defined on the original network $G$. We examine the fraction of infected target nodes for all methods and find that the targeted collective influence performs best as well (see Appendix \ref{othersimulation}).

We evaluate the ratio of infected non-target node to infected target node ($g(s_{n^*})/f(s_{n^*})$) in Fig. \ref{FIG:2}. A lower ratio indicates a better performance of the method. Here we only compare Eigs, HDs, CI and $\Delta CI$ as other methods rarely infect target nodes. Again, we find that the targeted collective influence outperforms other competing approaches and $\Delta CI_2$ performs better than $\Delta CI_1$.

\begin{figure}
	\centering
		\includegraphics[scale=.5]{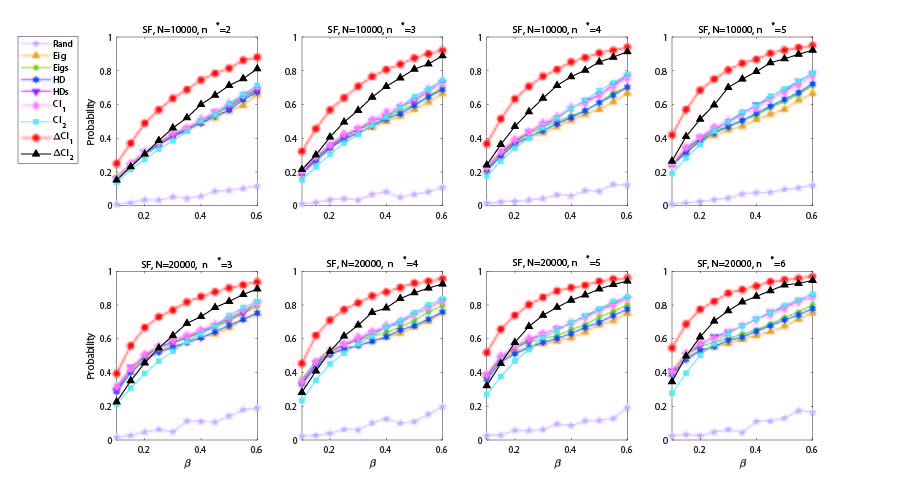}
	\caption{The probability that at least one target node is infected in scale-free (SF) networks with size $N=10,000$ or $20,000$ and degree distribution $P(k)\sim k^{-3}$. We use different methods (including Rand, Eig, Eigs, HD, HDs, CI and $\Delta CI$ with $\ell=1$ and $\ell=2$) to select the seeds for different size $n^{*}=2,\cdots,6$. The results are averaged over 100 independent realizations.}
	\label{FIG:3}
\end{figure}

\begin{figure}
	\centering
		\includegraphics[scale=.5]{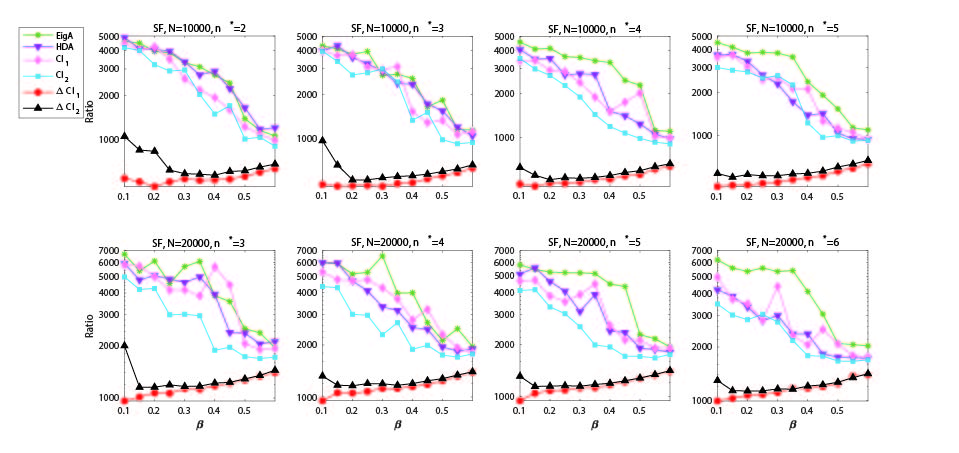}
	\caption{The ratio of $g(s_{n^*})$ to $f(s_{n^*})$ for scale-free (SF) networks with size $N=10,000$ or $20,000$. We compare the performances of different methods including Eigs, HDs, CI and $\Delta CI$ with $\ell=1$ and $\ell=2$. The results are averaged over 100 independent realizations.}
	\label{FIG:4}
\end{figure}

\begin{figure}
	\centering
		\includegraphics[scale=.45]{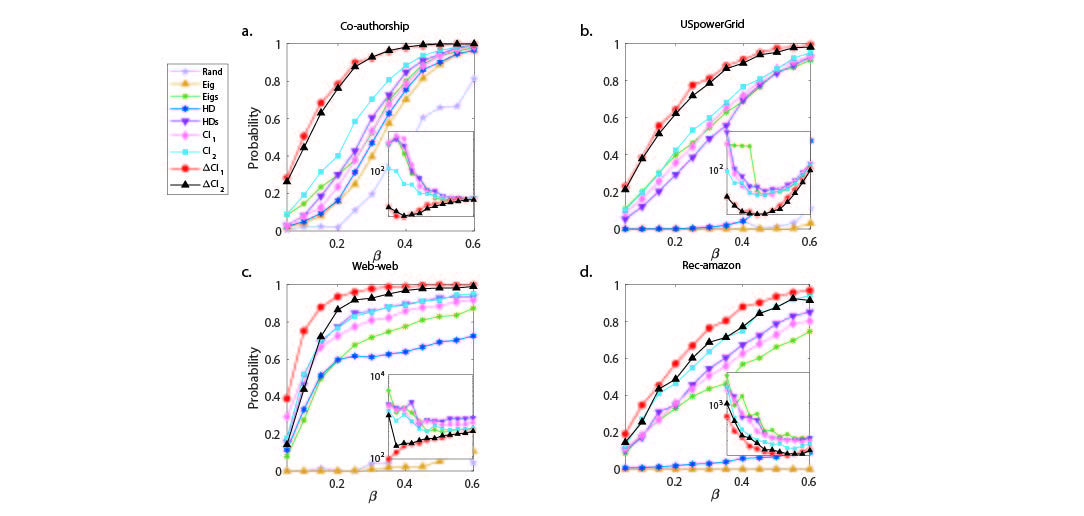}
	\caption{The probability that at least one target node is infected in four real networks for 10 target nodes. The inset in each panel shows the ratio of $g(s_{n^*})$ to $f(s_{n^*})$. We use different methods (including Rand, Eig, Eigs, HD, HDs, CI and $\Delta CI$ with $\ell=1$ and $\ell=2$) to select the seeds for size $n^{*}=4$. a, The co-authorship network of scientists with $N=379$ and $M=914$. b, The US power Grid network with $N=4.9K$ and $M=6.6K$. c, The web graph network with $N=16.1K$ and $M=25.6K$. d, The recommendation network of Amazon with $N=91.8K$ and $M=125.7K$.}
	\label{FIG:5}
\end{figure}

\subsection{Scale-free networks}

We perform the same analysis on scale-free (SF) networks with 10 target nodes. The SF networks have power-law degree distributions. We generate undirected SF networks with $N=10,000$ or $20,000$ using the preferential attachment model \cite{Albert1999Emergence}. Simulation results are shown in Figs. (\ref{FIG:3}) and (\ref{FIG:4}). We find that the targeted collective influence performs better than competing methods. Interestingly, for SF networks, $\Delta CI_1$ outperforms $\Delta CI_2$. This is possibly due to the existence of highly connected hubs. $\Delta CI_2$ may select global hubs that have both high $CI^{G_s}$ and $CI^{G_c}$ but are far from target nodes. In contrast, $\Delta CI_1$ can potentially select local hubs that are close to target nodes. Further analyses are needed to test this hypothesis in future works. 

\subsection{Real-world networks}

We finally validate the proposed method in several real-world networks \cite{rossi2015network}. Four real networks of distinct types are selected, including a co-authorship network ($N=379$ and $M=914$), the US power grid network ($N=4.9K$ and $M=6.6K$), a web graph network (links between webpages) ($N=16.1K$ and $M=25.6K$), and a recommendation network of Amazon ($N=91.8K$ and $M=125.7K$). Data sources and statistics of the networks are reported in Appendix \ref{networkdata}.

In the first set of experiments, we assign 10 target nodes in one cluster and aim to find $n^*=4$ seeds. Experiment results are shown in Fig. (\ref{FIG:5}). Consistent with simulations on synthetic networks, the targeted collective influence performs best. A same method can have different performance in the four real-world networks depending on the network structure. For instance, HD performs much worse in the US power grid network and the Amazon recommendation network. However, the good performance of $\Delta CI_1$ and $\Delta CI_2$ is robust across all tested networks. $\Delta CI_1$ is generally better than $\Delta CI_2$ in the four networks. For the more heterogeneous web graph network (the maximum degree is 1.7K and the average degree is 3), the advantage of $\Delta CI_1$ over $\Delta CI_2$ is more prominent, which agrees with the results in SF networks.

We further consider the case that target nodes are located in several clusters that spread across the network. Specifically, we select two clusters of target nodes, each cluster with 10 target nodes. This optimization problem is more challenging as target nodes are not located in one place. Results shown in Fig. (\ref{real20}) indicate that $\Delta CI_1$ and $\Delta CI_2$ still outperform competing methods. The advantage of $\Delta CI_1$ and $\Delta CI_2$ is more prominent in sparse networks with lower average degrees (e.g., the recommendation network of Amazon and the US power grid). We additionally test the case with 30 target nodes in three clusters. Results in Fig. (\ref{real30}) demonstrate the consistent better performance of $\Delta CI_1$ and $\Delta CI_2$.

\section{Conclusion}

Targeted influence maximization has broad applications in real-world problems. In this study, we formulated the SIR model using a message passing process, which can better represent the transmission dynamics, and further developed a theoretical framework to analyze the targeted influence maximization problem based on stability analysis and optimal perturbation. Our analysis led to the a metric, termed targeted collective influence, that was used to identify influential spreaders in targeted spreading process. We validated the proposed method in both synthetic and real-world networks, demonstrating its robust performance that out-competes commonly used approaches. Our study provides a theoretically based metric that was shown effective in a range of network structures.

\appendix
\section{Competing Methods}\label{othermethod}
We compare the targeted collective influence with several heuristic metrics that are widely used to rank the spreading capability of nodes. In numerical simulations, all nodes are ranked by each method and then the top $n^{*}$ nodes with highest scores are selected as the set of seeds $S_{n^{*}}$. 
\begin{itemize}
\item Eigenvector-based ranking. The dynamical importance of nodes can be quantified using the eigenvector corresponding to the largest eigenvalue of the adjacency matrix \cite{restrepo2006characterizing}. Using a perturbation analysis on the largest eigenvalue, the {\it dynamical importance} of a node $i$ is calculated as
\begin{equation}\label{eig}
I_{i}=\frac{v_i u_i}{\mathbf{v}^T \mathbf{u}},
\end{equation}
where $\mathbf{v}$ and $\mathbf{u}$ denote the right and left eigenvectors of the adjacency matrix $\{a_{ij}\}_{N\times N}$. In simulations, we use two versions of this method - Eig (for the original network) and Eigs (for the subgraph $G_{s}$).

\item Degree-based ranking. In high degree (HD) ranking, the score of each node $i$ is determined by the number of its connections: $K^{HD}_{i}=\sum_{j\in\partial i}a_{ij}$. We also compare with the HD ranking in the subgraph $G_{s}$ (HDs).

\item Collective influence. The collective influence (CI) of each node is computed using power iteration that aims to estimate the largest eigenvalue of the NB matrix of the network \cite{morone2015influence}. Specifically, the CI score of node $i$ at level $\ell$ is $CI_{\ell}(i)=\left(k_{i}-1\right)\sum_{j\in \partial Ball(i,2\ell-1)}\left(k_{j}-1\right)$.

\end{itemize}

\section{Additional experiments}\label{othersimulation}

\begin{figure}
	\centering
		\includegraphics[scale=.5]{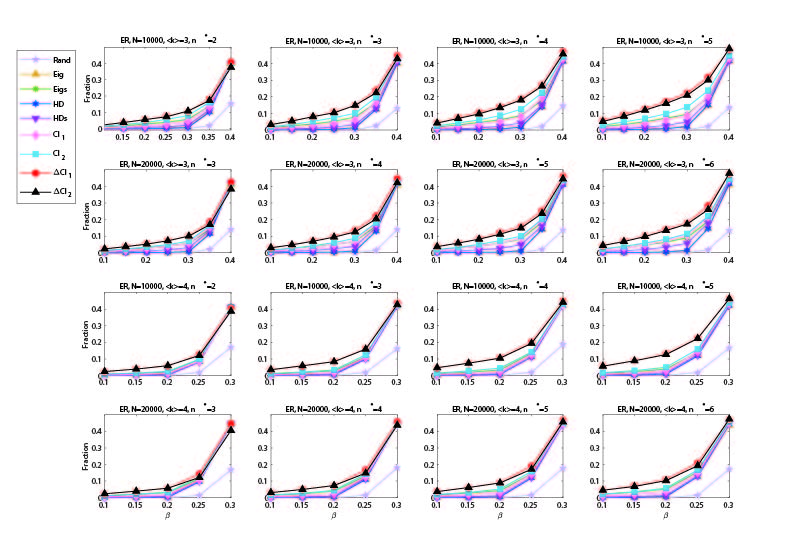}
	\caption{The fraction of infected target nodes in Erd\"{o}s-R\'{e}nyi (ER) networks with size $N=10,000$ or $20,000$ and mean degree $\left<k\right>=3$ or $4$. We use different methods (including Rand, Eig, Eigs, HD, HDs, CI and $\Delta CI$ with $\ell=1$ and $\ell=2$) to select seeds for different size $n^{*}=2,\cdots,6$. The results are averaged over 100 independent realizations.}
	\label{fracER}
\end{figure}

\begin{figure}
	\centering
		\includegraphics[scale=.5]{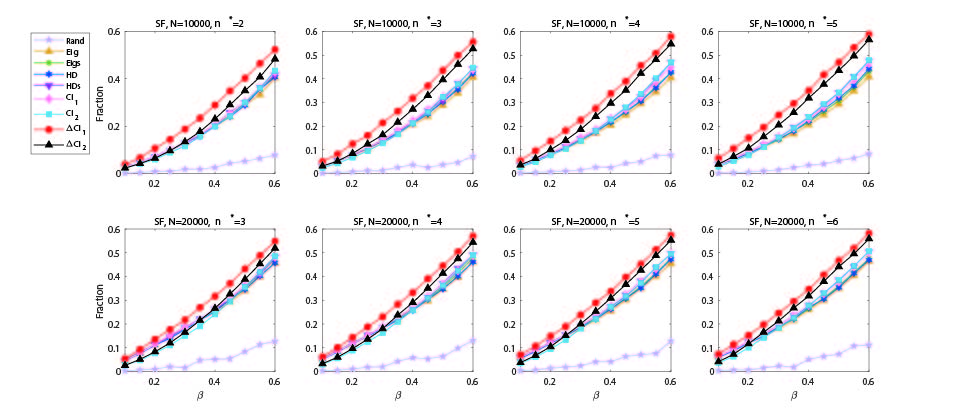}
	\caption{The fraction of infected target nodes in scale-free (SF) networks with size $N=10,000$ or $20,000$ and degree distribution $P(k)\sim k^{-3}$. We use different methods (including Rand, Eig, Eigs, HD, HDs, CI and $\Delta CI$ with $\ell=1$ and $\ell=2$) to select the seeds for different size $n^{*}=2,\cdots,6$. The results are averaged over 100 independent realizations.}
	\label{fracSF}
\end{figure}

\begin{figure}
	\centering
		\includegraphics[scale=.45]{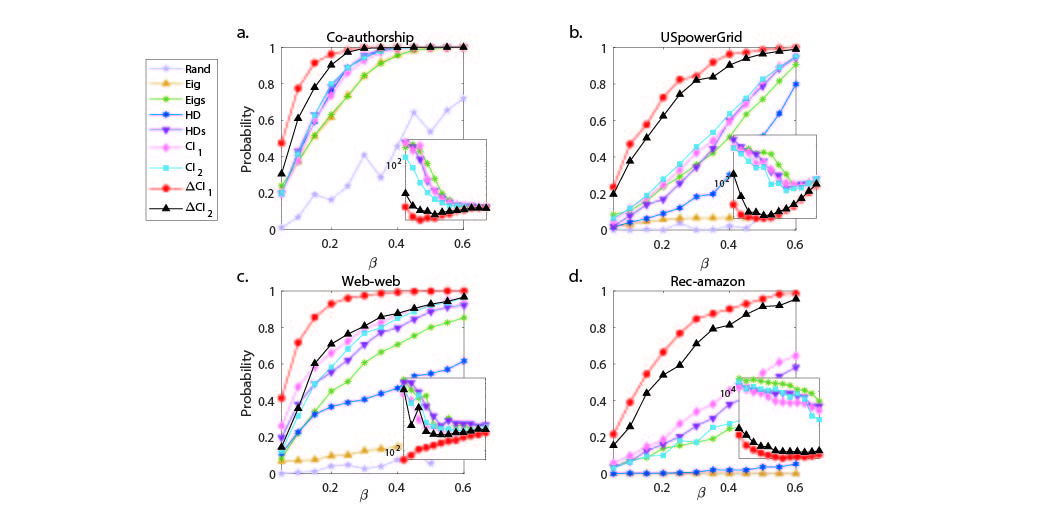}
	\caption{The probability that at least one target node is infected in four real networks with 20 target nodes. The inset in each panel shows the ratio of $g(s_{n^*})$ to $f(s_{n^*})$. We use different methods (including Rand, Eig, Eigs, HD, HDs, CI and $\Delta CI$ with $\ell=1$ and $\ell=2$) to select the seeds for size $n^{*}=4$. a, The co-authorship network of scientists with $N=379$ and $M=914$. b, The US power Grid network with $N=4.9K$ and $M=6.6K$. c, The web graph network with $N=16.1K$ and $M=25.6K$. d, The recommendation network of Amazon with $N=91.8K$ and $M=125.7K$.}
	\label{real20}
\end{figure}

\begin{figure}
	\centering
		\includegraphics[scale=.45]{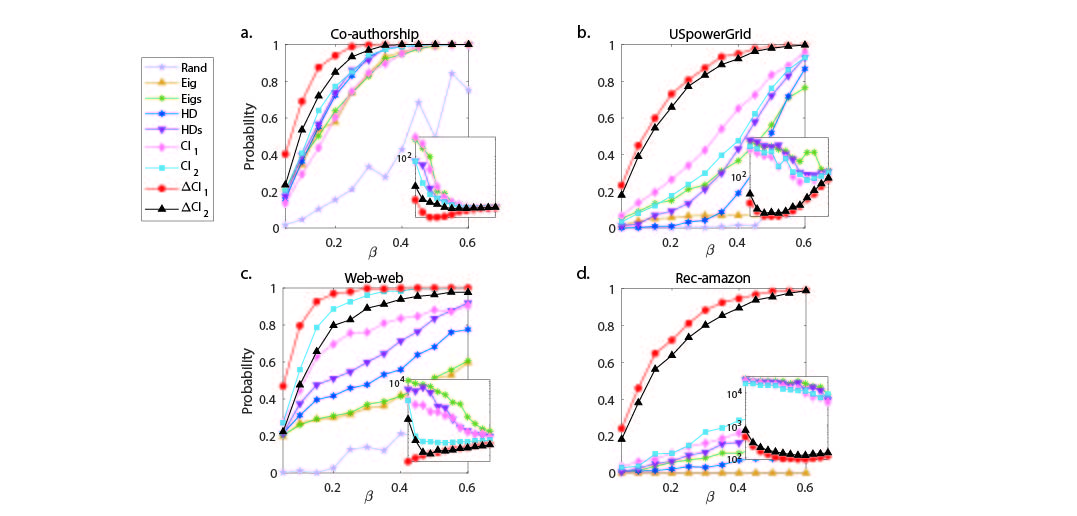}
	\caption{The probability that at least one target node is infected in four real networks with 30 target nodes. The inset in each panel shows the ratio of $g(s_{n^*})$ to $f(s_{n^*})$. We use different methods (including Rand, Eig, Eigs, HD, HDs, CI and $\Delta CI$ with $\ell=1$ and $\ell=2$) to select the seeds for size $n^{*}=4$. a, The co-authorship network of scientists with $N=379$ and $M=914$. b, The US power Grid network with $N=4.9K$ and $M=6.6K$. c, The web graph network with $N=16.1K$ and $M=25.6K$. d, The recommendation network of Amazon with $N=91.8K$ and $M=125.7K$.}
	\label{real30}
\end{figure}

Figures (\ref{fracER}) and (\ref{fracSF}) show the fraction of infected target nodes for all methods in Erd\"{o}s-R\'{e}nyi networks and scale-free networks respectively. Figures (\ref{real20}) and (\ref{real30}) show the results when target nodes are located in two and three clusters. 

\section{Network data} \label{networkdata}

Network data are downloaded from the following websites. (1) The co-authorship network of scientists (\url{https://networkrepository.com/ca-netscience.php}). (2) The US power Grid network (\url{https://networkrepository.com/USpowerGrid.php}). (3) The web graph network (\url{https://networkrepository.com/web-webbase-2001.php}). (4) The recommendation networks of Amazon (\url{https://networkrepository.com/rec-amazon.php}).

\begin{table}[width=.9\linewidth,cols=4,pos=h]
\caption{The properties of real-world networks used in this paper}\label{tbl1}
\begin{tabular*}{\tblwidth}{@{} LLLLLL@{} }
\toprule
& Size $N$ & Links $M$ & $\left<k\right>$ & Maximum Degree & Maximum k-core \\
\midrule
Co-authorship & 379 & 914 & 4 & 34 & 9 \\
US power Grid & 4.9K & 6.6K & 2 & 19 & 6 \\
Web-web graph & 16.1K & 25.6K & 3 & 1.7K & 33 \\
Rec-amazon & 91.8K & 125.7K & 2 & 5 & 5 \\
\bottomrule
\end{tabular*}
\end{table}

\printcredits

\section*{Acknowledgements}
\label{sec:Acknowledgments}

R.Zhang is supported by National Key Research and Development Program of China (Grant No.2021ZD0112400 and No.2020YFA0713702), National Natural Science Foundation of China (Grant No.11801058) and High-level Talents program of Dalian City (Grant No.2020RQ061).

\bibliographystyle{model1-num-names}

\bibliography{cas-refs}


\end{document}